\documentstyle[aps,preprint]{revtex}
\begin{document}

\title{String field theory Hamiltonians from Yang-Mills theories}
\author{Vipul Periwal}
\address{Department of Physics,
Princeton University,
Princeton, New Jersey 08544}

\def\DD{\hbox{D}}
\def\dd{\hbox{d}}
\def\tr{\hbox{tr}}\def\Tr{\hbox{Tr}}
\def\ee#1{{\rm e}^{{#1}}}
\def\part{\partial}
\def\bpart{\bar\partial}
\def\del#1#2{{{\delta #1}\over{\delta #2}}}
\def\refe#1{eq.~\ref{#1}\ }
\maketitle\tightenlines
\begin{abstract}
Marchesini showed that the Fokker--Planck Hamiltonian for Yang-Mills
theories is the loop operator.  Jevicki and Rodrigues 
showed that the 
Fokker--Planck Hamiltonian of some matrix models co\"\i ncides 
with temporal gauge non-critical string field theory Hamiltonians
constructed by Ishibashi and Kawai (and their collaborators).  
Thus the loop operator for Yang--Mills theory is the temporal
gauge Hamiltonian for a noncritical string field theory, in accord with
Polyakov's conjecture.
The consistency condition of the string interpretation is the zigzag 
symmetry emphasized by Polyakov. Several aspects of the noncritical 
string theory are considered, relating the string field theory 
Hamiltonian to the worldsheet description.
\end{abstract}
\bigskip
Mandelstam\cite{sm}\ realized the 
importance of gauge invariant loop observables in Yang--Mills theory.
See also \cite{others}.  
The fact that a string interpretation for
Yang--Mills theories is natural is particularly transparent 
in the loop equation derived by
Guerra and collaborators\cite{guerra}, and 
rediscovered in \cite{migdal}. This Schwinger--Dyson equation governs the
dynamics of gauge invariant Wilson loops in Yang--Mills theory.
There are 
simple geometric interpretations for the various terms that appear 
in terms of string propagation and interactions, with the string joining 
interaction suppressed by a factor of $1/N^{2}$ relative to the string 
splitting interaction  in a normalization natural for the large $N$
limit.   

The loop equation governing Wilson loop expectation values in gauge theories 
can be written as the expectation value of the 
action of an operator, the loop operator, acting
on Wilson lines.  It is a fundamental observation due to Marchesini\cite{marc}
that the loop operator co\"\i ncides with the Fokker--Planck 
Hamiltonian that appears in the stochastic quantization\cite{parisi}\ 
of the gauge theory.  As such, the loop operator plays a central 
r\^ole in gauge theories but its significance in, for example, a
string field theory equivalent to a Yang--Mills theory has not been
elucidated.  The contribution of this paper is to point out that {\it the
loop operator is precisely the temporal gauge string field theory 
Hamiltonian.} I show that the consistency condition for this 
identification is Polyakov's zig--zag symmetry.
I  explain how the peculiar asymmetry between
joining and splitting vertices can be  accounted for in a worldsheet 
description, and I  explain how recent conjectures on Yang--Mills
theories and their dimensional reductions\cite{pol,juan,review}
are related to  the explicit identification given here.  

The explicit connection given in this paper between
string theories and gauge theories should be compared to 
the efforts that have gone into 
attempted constructions of worldsheet descriptions of 
first--quantized string propagation in supergravity 
backgrounds\cite{efforts}\ 
believed to be dual to gauge theories. The conceptual simplicity of the 
string field theory Hamiltonian is   striking 
because in no other instance in string theory is a string field 
theory simpler or easier to construct than the first--quantized string theory.  
The identification of this  Hamiltonian can be used to { deduce} 
properties of a worldsheet first--quantized description, as 
I shall show below.  It is plausible that the first--quantized 
description will capture only part of the physics incorporated in the
{\it exact} string field theory Hamiltonian, just as an expansion in 
Feynman diagrams misses non--perturbative physics.  

Concretely, the string field theory that arises from the Yang--Mills 
theory is a {\it noncritical} string theory.  The spacetime 
dimension in which the string propagate is therefore one more than the 
sum of the number of adjoint scalars and the number of gauge 
potential components.  Thus, {\it e.g.} for the $d=4$ maximally supersymmetric 
Yang--Mills theory, the spacetime dimension of the noncritical 
string theory is 11, which differs from the  na\"\i ve interpretation 
of the Maldacena conjecture\cite{juan}.   However the present 
construction precisely agrees
with the conjectured QCD string in Polyakov's approach\cite{pol} which
has been argued on physical grounds to be a noncritical string 
living in five dimensions.  Since the infinite $N$ Maldacena 
conjecture is supported by several precise comparisons with 
IIB supergravity in ten dimensions, 
it is important to explain how a noncritical string
theory can maintain this agreement.  I explain this below by 
recalling the difference between noncritical string measures and
critical string measures following \cite{rob,kkp}.
Furthermore, comparing to the calculation of
Dorey, Hollowood, Khoze, Mattis and Vandoren\cite{mattis}, it is interesting to 
observe that at finite $N$ the instanton calculation does not localize
on $S^{5}$ but rather seems to involve an 11--dimensional 
space\cite{grosspot}.  (The connection to the noncritical 
string given in a previous version of this paper 
was not mentioned in \cite{grosspot}.)

The loop operator is easiest understood in
the lattice theory since a precise interpretation requires a 
cutoff---this cutoff is not necessarily the same as the cutoff used in 
defining the quantum gauge theory\cite{sashabook}.
Define the loop 
function for a U$(N)$ gauge theory on a $d$--dimensional lattice 
\begin{equation}
	W(C) \equiv {1\over N} \Tr U(C)
\end{equation}
where $C$ is a closed contour on the lattice, and $U(C)$ is the path ordered
product of link variables along this contour 
$U_{\mu_{1}}(x_{1})U_{\mu_{2}}(x_{2})\dots U_{\mu_{n}}(x_{n}).$  
Define an electric field operator appropriate for a $d+1$--dimensional
theory\cite{marc}:
\begin{equation}
	E_{\mu}\ : \ \left[E_{\mu}^{a}(x), U_{\nu}(y)\right] = T^{a}U_{\nu}(y) 
	\delta_{\mu\nu}\delta_{x,y}
\end{equation}
where $\mu,\nu=1,\dots d,$ and $a=1,\dots N^{2},$ with $\Tr 
T^{a}T^{b} = N/2.$
For any lattice action, $S(U),$ the stochastic 
Fokker--Planck Hamiltonian is
\begin{equation}
	H \equiv {2\over N} \sum_{x,\mu,a} \ee{-S(U)} E_{\mu}^{a}(x) \ee{S(U)}
	E_{\mu}^{a}(x) \ ;
\end{equation}
while $H$  is not Hermitian, by a similarity 
transformation we can make it Hermitian.

There are simple geometric interpretations for 
the terms that appear in $H,$ which are clearer when we write 
\begin{equation}
	H \equiv {2\over N} \sum_{x,\mu,a}   E_{\mu}^{a}(x)  
	E_{\mu}^{a}(x) + \left[ E_{\mu}^{a}(x), {S(U)}\right]
	E_{\mu}^{a}(x). 
\end{equation}
The first term, which is independent of the
lattice action, includes terms that correspond to the splitting and 
joining of strings, with the latter term suppressed relative to the
former by a factor of 
$N^{-2}.$  The equilibrium condition for the stochastic
field theory correlation function turns out to be the loop 
equation\cite{marc}:
\begin{equation}
<HW(C)> = 0. 
\end{equation} 
Independently, Jevicki and Sakita\cite{coll}\ showed 
that the large $N$ saddle point equation of Yang--Mills theory 
formulated as a collective field theory  is  equivalent to 
the loop equation, as well.

The asymmetry in the powers of $N$ for joining and splitting 
interactions is a clue for  
the string field theory interpretation since in temporal gauge 
one does find such an asymmetry between string splitting and string
joining vertices\cite{ik}.  This reassignment  of
powers of $N$ is consistent with the usual topological 
loop counting of string diagrams.
The $S$--dependent term has operators corresponding to the
motion in the space of loops, and tadpole operators that annihilate loops.  
Thinking of $S$ as defining the background configuration of the space 
of loops, the propagation and annihilation of loops is background 
dependent, but the interactions of loops are background independent.
Notice that the `t Hooft coupling $g_{\rm YM}^{2}N$ appears only in 
this term.
This is essentially another realization of an idea of Horowitz, 
Lykken, Rohm and Strominger\cite{cubic} regarding string field 
theory.   
In the large $N$ limit there are no terms that can join strings.  Thus
in the large $N$ limit, $H$ acts as a derivation on products 
$\prod_{i}W(C_{i}).$

In a beautiful set of papers, Ishibashi, Kawai, and 
collaborators\cite{ik,kawaitemp,ik2}\ 
have constructed non-critical string field theories in temporal gauge.  
As an example, consider 
their string field theory Hamiltonian for the $c=0$ model:
\begin{equation}
	H_{\rm IK} = \int_{0}^{\infty} \dd l_{1}\dd l_{2} \Psi^{\dagger}_{l_{1}}
\Psi^{\dagger}_{l_{2}}\Psi_{l_{1}+l_{2}}(l_{1}+l_{2}) +
\int_{0}^{\infty} \dd l \rho(l) \Psi_{l} + O(g^{2}) 
\end{equation}
where $\Psi_{l}$ annihilates a string of length $l,$ and 
$\left[\Psi_{l},\Psi^{\dagger}_{l'}\right] = \delta(l-l').$  The term 
of $O(g^{2})$ is  
\begin{equation}
	g^{2} \int_{0}^{\infty} \dd l_{1}\dd l_{2} \Psi^{\dagger}_{l_{1}+l_{2}}
\Psi_{l_{2}}\Psi_{l_{1}}l_{1}l_{2}
\end{equation}
and describes the merging of strings.  $\Psi$ annihilates unmarked 
loops and $\Psi^{\dagger}$ creates marked loops.  The string joining  
and string splitting terms have different powers of $g,$ just as 
there are different powers of $N^{{-1}}$ in the loop joining and splitting 
terms in the loop operator.  This Hamiltonian was first written down for 
the $c=1$ model by Das and Jevicki\cite{das}\ and it was pointed out
subsequently by Moore, Seiberg and Staudacher\cite{moore}\ 
that it applied as well 
to $c<1$ models with changes in the tadpole term.  

A connected amplitude with $b$ 
boundaries and $h$ handles comes with a factor $g^{2h-2+2b},$ which 
is not the usual topological combination.  (I have corrected 
a small error in \cite{ik2}\ here.) This can be traced to the fact that 
the disk amplitude is non-zero and has a factor $g^{0}$ in this 
formalism.  Normalizing connected amplitudes with respect to the
conventionally normalized disk amplitudes gives the usual Euler 
characteristic power $g^{-(2-2h-b)}.$  In the temporal gauge 
interpretation, the tadpole term is 
$O(1),$ just as for large $N$ Yang--Mills theory the  
normalized expectation value is $O(1).$  One could rescale $\Psi$ and
$\Psi^{\dagger}$ while preserving the commutation relation to make
the joining and the splitting terms of the same order in $g,$ but this
would change the normalization of the tadpole term as well, making the
expectation value of a Wilson loop $O(N),$ appropriate for a string 
disk amplitude.  This is not, however, natural from the point of view 
of the gauge theory, so I will give below another geometric 
interpretation of the powers of $g.$

Jevicki and Rodrigues\cite{jev2} showed that the Hamiltonian 
$H_{\rm IK}$ (including the $O(g^{2})$ 
term)   arises naturally as the double--scaling limit\cite{matrix}\ of the
$k=2$ matrix model Fokker--Planck Hamiltonian.  Thus one 
has a direct map from matrix models to noncritical 
string field theories in the
temporal gauge, via the Fokker--Planck Hamiltonian.  

Crucial to the string interpretation of the Fokker--Planck 
Hamiltonian is a consistency check, 
equivalent to diffeomorphism invariance on the string 
worldsheet\cite{ik2,jev2}.  In the case of the loop operator, it is a 
generalization of the derivation given in \cite{ik2}(Sect.~3) that the
consistency of the string interpretation is equivalent to the 
zigzag symmetry of Wilson loops, particularly emphasized by 
Polyakov\cite{pol}.  The key point to note here is that two Wilson 
loops $L_{i}, i=1,2$ (which are the strings in this formalism) may join 
either by an
infinitesimal loop attaching to $L_{1}$ leading to contact with $L_{2}$
or vice versa.  For the string interpretation, diffeomorphism 
invariance on the worldsheet implies that the difference between these
two amplitudes should vanish.  Subtracting the amplitudes for the two
processes, we encounter amplitudes involving insertions of 
infinitesimal back--tracking loops, which are trivial if and only if 
the zig--zag symmetry holds.

In connection with the zig--zag symmetry,  a
recent investigation  of the loop equation\cite{review} in the AdeS/CFT 
correspondence\cite{gross}\ has found that
the agreement between gauge theory expectations and critical string 
expectations is regularization dependent.  Recalling the analysis of
Polyakov\cite{pol} and Klebanov, Kogan and Polyakov\cite{kkp}\ suggests then
that the zig--zag symmetry requires the non--critical Liouville 
dimension to be taken into account.

Thus we reach the main conclusion of this paper:  The loop operator of 
Yang--Mills theory 
is the temporal gauge string field theory Hamiltonian of a 
noncritical string
theory, provided that the zig--zag symmetry is maintained. 

This observation suggests that the appropriate place to look for a 
worldsheet description of the Yang--Mills string is temporal 
gauge\cite{kawaitemp}.  This gauge is quite different from conformal 
gauge so much of the usual intuition for first--quantized string 
theory requires revision.  For example,   
in light--cone string field theory, for example, string splitting and 
joining interaction vertices are given by mirror image worldsheets, with 
curvature singularities precisely at the joining/splitting point.  
These curvature singularities are integrable and precisely lead to the
expected power of $g$ for both string joining and splitting vertices, 
since the spacetime dilaton $\Phi$ couples to $\int {}^{(2)}R.$
The incoming and outgoing strings lie on curves with vanishing geodesic
curvature in light--cone string field theory.  In temporal gauge, as
mentioned above, the splitting and joining interactions come with 
different powers of $g.$  Since the Euler characteristic of the pants
diagram is $\chi_{E}=-1,$ how can this be?  A simple resolution of this puzzle 
is to note that only the worldsheet bulk curvature couples to the 
dilaton, whereas the Euler characteristic is a sum   of two terms
\begin{equation}
\int {}^{(2)}R + \sum_{\rm boundaries}\oint_{C_{i}}
\kappa = 2\pi\chi_{E}
\end{equation}
according to
Gauss and Bonnet\cite{bonnet}, with $\kappa$ the geodesic curvature of
the boundary components.
There is a time asymmetry in the definition of temporal gauge, so we
can sensibly distinguish between incoming and outgoing strings.  
Let $\oint \kappa = 2\pi K_{\pm}$ for `standard' 
outgoing/incoming loops respectively, then for the joining vertex we have
\begin{equation} 
{1\over 2\pi}\int {}^{(2)}R = -2 = \chi_{E}(\hbox{pants}) - K_{+} - 2K_{-}
\end{equation}
and for the splitting vertex we have
\begin{equation} 
{1\over 2\pi}\int {}^{(2)}R = 0 = \chi_{E}(\hbox{pants}) - 2K_{+} - K_{-}\ .
\end{equation}
We find then
\begin{equation}
K_{+}=-K_{-}=-1.
\end{equation}
This is {\it precisely} as it must be
since in the standard gluing of worldsheet diagrams in string field 
theory, the geodesic curvatures of boundaries being identified must 
cancel for constructing smooth surfaces.  As an 
independent check this predicts that the tadpole diagram (a disk) 
will have one incoming string giving a contribution $K_{-}=+1$ to the Euler 
characteristic and therefore have vanishing integrated bulk curvature,
\begin{equation}
\chi_{E}(\hbox{disk}) - K_{-} \ = +1 - 1 = 0 = {1\over 2\pi}\int {}^{(2)}R 
\end{equation}
implying that the tadpole will be $O(1)$, which is exactly correct.
As another consistency check,  
the Euler characteristic of the cylinder is zero,thus
incoming and outgoing boundaries must have boundary geodesic curvature 
that differs only in sign since the propagation term in the 
loop operator has no factors of $N$ and therefore corresponds to a 
vanishing bulk contribution.
The fact  that there is nonvanishing geodesic 
curvature associated with the boundaries in this gauge is related
to the fact that the Wilson loops are normalized with factors of $1/N.$

The discussion trivially extends to dimensionally reduced and/or 
supersymmetric\cite{itoyama}\ 
gauge theories, with $A_{\mu}(x(s)){\rm d}{x^{\mu}}/{\rm d}s$
replaced by $\Phi_{i}
(x(s))\sum_{n}v^{i}\delta({s-s_{n}})$ in the path--ordered 
exponential for dimensionally reduced directions.  
This is of interest in light of 
the conjectures of Banks, Fischler, Shenker and Susskind\cite{bfss},
and Maldacena\cite{juan}, though it needs to emphasized again that 
these conjectures are associated with critical string theories.   
 
In Polyakov's attempts at constructing
string theories equivalent to Yang--Mills theory, on the other hand,
the strings that appear are non--critical strings\cite{pol}.  As such, the 
identification which has been proposed in this paper is exactly in line
with his work, especially noting the zig--zag symmetry's r\^ole in a
string interpretation.  

The string field theory that I have 
identified in this paper is gauge--fixed.  This  is related to
the fact that we have worked entirely in terms of  gauge--invariant
operations on gauge--invariant Wilson loops. Recollect that residual  
spacetime gauge transformations in critical string theory  appear as 
conformal transformations on the worldsheet in conformal gauge.

Since I have given  here a construction of a non--critical string theory from 
Yang--Mills theory that is different from the standard 
conjectures\cite{review}, I must explain how the numerous agreements
supporting the standard conjectures 
that appear at string tree--level come about.  This is easily done:
What is the difference between a critical
string amplitude and a noncritical string amplitude?  The functional measure 
for a critical string amplitude has an integration over the conformal 
mode and a division by the volume of
the group of conformal transformations, vol({\it Conf}).  
These cancel for correlation 
functions of on--shell vertex operators, and do not cancel if we
consider off--shell amplitudes\cite{rob}.  Even for off--shell
amplitudes, on the sphere, there are no 
moduli and vol({\it Conf}) is a amplitude independent constant.  
On higher genus surfaces, vol({\it Conf}) 
depends on the moduli and cannot be factored out.  
For noncritical strings, there is still an integration 
over the conformal mode, but no corresponding division by vol({\it Conf}).
Thus, if we are interested in tree amplitudes, we can extract 
critical string tree amplitudes from a subset of the 
noncritical string tree amplitudes.  (Recall that string tree amplitudes do not 
suffice to reconstruct string loop amplitudes\cite{mark}.)
This subset corresponds to 
on--shell string states that have no dependence on the conformal mode.   
From the Fokker--Planck perspective, it is natural to focus 
on correlations 
\begin{equation}
\lim_{D\uparrow\infty}\langle 0| \exp(-DH) \prod W(C_{i})|0\rangle 
\end{equation}
where $D$ is the Fokker--Planck time (the Liouville dimension 
co\"ordinate), here identified with the worldsheet time co\"ordinate, 
of Wilson loops $C_{i}$ that satisfy the equilibrium condition
\begin{equation}
	\left[H,W(C_{i})\right] = 0 \ \hbox{for each } i.
\end{equation}
It  seems clear that 
correlations of observables satisfying 
the equilibrium condition will correspond  to amplitudes with
truncated external leg
propagators, natural for on--shell vertex operators in critical string theory.
However,  the precise relation between 
these two sets of objects remains to be computed.

Lastly, I mention that supporting evidence for my identification 
comes from a very recent preprint by Lidsey\cite{lid}.  In this work
it is shown that the critical IIB supergravity backgrounds  can be embedded 
in eleven--dimensional Ricci--flat spaces, including backgrounds with
non--trivial Ramond--Ramond fields.  This is precisely in accord with
the above--mentioned relation between critical strings and 
non--critical strings at tree level, see especially Klebanov, Kogan 
and Polyakov\cite{kkp}.

I am grateful to  I. Klebanov, G. Lifschytz,  J. Polchinski and
A. Polyakov 
for helpful conversations and A. Jevicki for a useful communication. 
This work was supported in part by NSF grant PHY-9802484.
\def\np#1#2#3{Nucl. Phys. B#1 (#2) #3}
\def\prd#1#2#3{Phys. Rev. D#1 (#2) #3}
\def\prl#1#2#3{Phys. Rev. Lett. #1 (#2) #3}
\def\pl#1#2#3{Phys. Lett. B#1 (#2) #3}


\begin{thebibliography}{99}
	
\bibitem{sm} S. Mandelstam, Ann. of Phys. 19 (1962) 25; Phys. Rev. 
175 (1968) 1580

\bibitem{others} J.--L. Gervais and A. Neveu, 
\pl{80}{1979}{255}; Y. Nambu, \pl{80}{1979}{372}; A. Polyakov, 
\pl{82}{1979}{247} 

\bibitem{guerra} G. De Angelis, D. De Falco and F. Guerra, Nuovo Cim. 
Lett. 19 (1977) 55; F. Guerra, R. Marra and G. Immirzi, Nuovo Cim. 
Lett. 23 (1978) 237

\bibitem{migdal} D. Foerster, Phys. Lett. 87B (1979) 83; T. Eguchi,
Phys. Lett. 87B (1979) 91; Yu. Makeenko 
and A. Migdal, Phys. Lett. 88B (1979) 135

\bibitem{marc} G. Marchesini, Nucl. Phys. B191 (1981) 214; B239 (1984) 
135

\bibitem{parisi} G. Parisi and Y.-S. Wu, Sci. Sin. 24 (1981) 484

\bibitem{pol} A. Polyakov, Nucl. Phys. Proc. Supp. 68 (1998) 1;
Int. J. Mod. Phys. A14 (1999) 645

\bibitem{juan} J. Maldacena, Adv. Theor. Math. Phys. 2 (1998) 231;
S. Gubser, I. Klebanov and A. Polyakov, \pl{428}{1998}{105}; E. 
Witten, Adv. Theor. Math. Phys. 2 (1998) 253

\bibitem{review} O. Aharony, S. Gubser, J. Maldacena, H. Ooguri 
and Y. Oz, {\sl Large $N$ field theories, string theory and gravity},
hep-th/9905111
\bibitem{rob} R.C. Myers and V. Periwal, \prl{70}{1993}{2841}
\bibitem{kkp} I.R. Klebanov, I. Kogan and A.M. Polyakov, 
\prl{71}{1993}{3243}
\bibitem{efforts} Reviews of this work can be found in the talks of
N. Berkovits and  R. Kallosh at Strings '99 (July 1999, Potsdam), 
available at http://strings99.aei-potsdam.mpg.de/cgi-bin/viewit.cgi



\bibitem{mattis} N. Dorey, T. Hollowood, V. Khoze, M. Mattis and S. 
Vandoren, \np{552}{1999}{88}

\bibitem{grosspot} D. Gross, talk at Strings '99 (July 1999, Potsdam), 
available at http://strings99.aei-potsdam.mpg.de/cgi-bin/viewit.cgi
 

\bibitem{sashabook} See {\it e.g.} A.M. Polyakov, {\sl Gauge fields 
and strings}, Harwood Academic Publishers (Chur, Switzerland, 1987)

\bibitem{coll} A. Jevicki and B. Sakita, Nucl. Phys. B185 (1981) 89

\bibitem{ik} N. Ishibashi and H. Kawai, \pl{314}{1993}{190}

\bibitem{cubic} G. Horowitz, J. Lykken, R. Rohm and A. Strominger,
\prl{57}{1986}{283}

\bibitem{kawaitemp} M. Fukuma, N. Ishibashi, H. Kawai and M. Ninomiya,
\np{427}{1994}{139}; M. Ikehara, N. Ishibashi, H. Kawai, T. Mogami, R. 
Nakayama and N. Sasakura, \prd{50}{1994}{7467}


\bibitem{ik2}  N. Ishibashi and H. Kawai, \pl{322}{1994}{67}; 
\pl{352}{1995}{75}



\bibitem{das} S. Das and A. Jevicki, Mod. Phys. Lett. A5 (1990) 1639

\bibitem{moore} G. Moore, N. Seiberg and M. Staudacher, \np{362}{1991} 
665

\bibitem{jev2} A. Jevicki and J. Rodrigues, \np{421}{278}{1994}
\bibitem{matrix} E. Brezin and V. Kazakov,  \pl{236}{1990}{144}; 
M. Douglas and S. Shenker, \np{335}{1990}{635}; D. Gross and A. Migdal, 
\prl{64}{1990}{127} 
\bibitem{gross} N. Drukker, D. Gross and H. Ooguri, {\sl Wilson loops 
and minimal surfaces}, hep-th/9904191
 

\bibitem{bonnet} O. Bonnet, J. \'Ecole Polytech. 19 cah. 32 (1848) 1

\bibitem{itoyama} For a recent comprehensive treatment see 
H. Itoyama and H. Takashino, Prog. Theor. Phys. 97 (1997) 963

\bibitem{bfss} T. Banks, W. Fischler, S. Shenker and L. Susskind, 
\prd{55}{1997}{5112}



 
\bibitem{mark} M. Srednicki and R. Woodard, \np{293}{1987}{612}
 
\bibitem{lid} J.E. Lidsey, {\sl The embedding of superstring 
backgrounds in Einstein gravity}, hep-th/9907095

 
 
 

\end{thebibliography}
\end{document}